\documentclass[11pt]{article}
\usepackage{amsmath,amssymb,color,graphics,epsfig,cite}

\usepackage{amsfonts}

\newcommand{\hoch}[1]{$\, ^{#1}$}

\newcommand{\be}{\begin{equation}}
\newcommand{\ee}{\end{equation}}
\newcommand{\bea}{\setlength\arraycolsep{2pt} \begin{eqnarray}}
\newcommand{\eea}{\end{eqnarray}}
\newcommand{\nn}{\nonumber}

\def\ft#1#2{{\textstyle{\frac{\scriptstyle #1}{\scriptstyle #2} } }}
\def\fft#1#2{{\frac{#1}{#2}}}

\def\0{{\sst{(0)}}}
\def\1{{\sst{(1)}}}
\def\2{{\sst{(2)}}}
\def\3{{\sst{(3)}}}
\def\4{{\sst{(4)}}}
\def\5{{\sst{(5)}}}
\def\6{{\sst{(6)}}}
\def\7{{\sst{(7)}}}
\def\8{{\sst{(8)}}}
\def\sst#1{{\scriptscriptstyle #1}}
\def\oneone{\rlap 1\mkern4mu{\rm l}}

\begin{document}

\begin{center}
{\Large {\bf New BPS States from Bosonic/Heterotic Duality }}

\vspace{20pt}

Kai-Peng Lu\hoch{1}, H. L\"{u}\hoch{1,2} and Liang Ma\hoch{1}
		
\vspace{10pt}
		
{\it \hoch{1}Center for Joint Quantum Studies, Department of Physics,\\
			School of Science, Tianjin University, Tianjin 300350, China }

\medskip
		
{\it \hoch{2}The International Joint Institute of Tianjin University, Fuzhou,\\ Tianjin University, Tianjin 300350, China}
		
\vspace{40pt}

\underline{ABSTRACT}
\end{center}

In this paper, we consider the recently proposed bosonic/heterotic duality that relates the heterotic superstrings to the noncritical bosonic string. Although the latter is nonsupersymmetric, it can be viewed as pseudo-supersymmetric in that the theory admits a consistent set of Killing spinor equations whose integrability conditions are satisfied by the equations of motion. We construct two classes of the dual pairs of BPS and pseudo-BPS solutions in the respective heterotic and bosonic strings and obtain their explicit Killing spinors. The existence of such dual pairs lends support to the proposed duality.

\vfill {\footnotesize kaipenglu@tju.edu.cn \ \ \  mrhonglu@gmail.com\ \ \ maliang0@tju.edu.cn}


\thispagestyle{empty}
\pagebreak



\section{Introduction}

Bogomol’nyi-Prasad-Sommerfield (BPS) or supersymmetric solutions in supergravity play an important role in the study of string theories, owing to the fact that some properties derived from these solutions, such as the mass/charge relation, may survive the quantum correction. In the early days of constructing BPS solutions, one typically considered solitonic configurations that describe $p$-branes, such as supermembranes \cite{Hughes:1986fa,Bergshoeff:1987cm} or D-branes \cite{Dai:1989ua,Polchinski:1995mt,Bergshoeff:1996tu}. Notable examples of supergravity solutions include the M-branes \cite{Duff:1990xz,Gueven:1992hh} and D3-branes \cite{Horowitz:1991cd,Duff:1991pea}. These solutions are charged under higher-form fields and are a generalization of the Reissner-Nordstr\"om (RN) black holes. In the extremal limit, the mass and charges are linearly related. One can obtain the Killing spinors of these bosonic backgrounds and demonstrate that they preserve a certain fraction of supersymmetry. Classifications of BPS $p$-brane solutions in various supergravities can be found in \cite{Duff:1994an,Bergshoeff:1996rn,Lu:1997hb}.

The construction of the BPS solutions becomes more sophisticated later, especially in lower dimensions where the supersymmetry becomes relatively simpler. The assumption of the existence of Killing spinor may allow one to construct all possible BPS solutions \cite{Gauntlett:2002nw,Gutowski:2003rg}. In this paper, we adopt a quite different approach. We construct new BPS solutions by making use of the previously proposed duality between the noncritical bosonic and heterotic strings, and we obtain two classes of BPS solutions in the heterotic theory. Indeed, it was shown \cite{Ma:2025mvo} that one can obtain the warped Mink$_4\times \mathbb R\times T^{1,1}$ vacuum of the heterotic theory from the Mink$_4\times S^3\times S^3$ vacuum of the noncritical bosonic string. Furthermore, the new heterotic vacuum is supersymmetric and preserving $1/4$ of the supersymmetry, even though the solution is motivated by the bosonic string.

At first sight, this is puzzling and one might attribute this emergence of supersymmetry to pure luck. However, mapping a solution of a bosonic string to a BPS one in superstring may not be farfetched. The bosonic string is special in that the theory in general dimensions all admits Killing spinor equations whose integrability conditions are satisfied precisely by the full set of equations of motion \cite{Lu:2011zx,Liu:2012jra,Lu:2011ku}. This is analogous to the situation where Einstein gravity in general dimensions admits the Killing spinor equation that is consistent with the Einstein field equation. This property allows one to construct pseudo-supergravity extension of the bosonic string, where up to the quadratic order in fermions, the Lagrangian is invariant under the pseudo-supersymmetric transformation rules derived from the Killing spinor equations \cite{Lu:2011ku}. Thus, if one establishes a duality in the bosonic sector between the heterotic theory and bosonic string, their Killing spinor equations, which are linear in fermions, may also survive the duality. This provides a new method of constructing BPS solutions of the heterotic theory. Of course, one can in principle construct these BPS solutions directly in the heterotic theory; however, it turns out that the bosonic dual solutions typically appear to be simpler and easier to construct.

The work of \cite{Ma:2025mvo} focused on the properties of the vacuum pair of the bosonic and heterotic strings. In this paper, we shall construct more examples of dual pairs of BPS  and pseudo-BPS solutions in the respective heterotic and bosonic strings. We shall not only construct the Killing spinors in the heterotic theory, but also in the noncritical bosonic string. The existence of such dual pairs provide supporting evidence for the proposed duality.

The paper is organized as follows. In Section 2, we give a review of the bosonic/heterotic duality, and provide some new results concerning the explicit Killing spinors on the Mink$_4\times S^3\times S^3$ vacuum of the noncritical string. In Section 3, we obtain $D=4$, ${\cal N}=1$ supergravity with a tensor supermultiplet from either heterotic supergravity or noncritical bosonic string. In four dimensions, the Hodge dual implies that the tensor multiplet is equivalent to complex scalar multiplet. We construct both BPS cosmic string and instanton solutions and obtain their Killing spinors.  In Section 4, we lift the $D=4$ solutions back to $D=10$ heterotic and noncritical bosonic strings and obtain their respective Killing spinors. The existence of the Killing spinor implies that the new cosmic strings and instantons are BPS solutions in the heterotic theory.  We conclude the paper in Section 5.  In order to construct explicit Killing spinors, it is necessary to obtain the spin connections for some natural choices of the vielbein for the metrics of our solutions. In Appendix A, we list all the spin connections. Since the round $S^3$ appears frequently in our solutions, we construct its Killing spinors in Appendix B, treating $S^3$ as a $U(1)$ bundle over $S^2$, and provide results in two different vielbein bases.

\section{Bosonic/Heterotic string duality}

\subsection{The mapping}

In \cite{Ma:2025mvo}, an intriguing duality map was proposed between the ten-dimensional noncritical bosonic string with a conformal anomaly term and the bosonic sector of the heterotic string. In general, there exist two heterotic string theories, one with $E_8\times E_8$ Yang-Mills fields, whilst the other has $SO(32)$. The simplest realization of the proposed duality involves only a $U(1)$ matter field. The relevant Lagrangian is
\bea
\hat{\mathcal{L}}_{10}&=& \hat{R}\, {\hat *\oneone}-\ft12 {\hat * d\hat \phi}\wedge d\hat \phi -\ft12 e^{-\hat\phi} {\hat *\hat H_\3} \wedge \hat H_\3 -\ft{1}{2 }e^{-\frac{1}{2}\hat{\phi}}{\hat *\hat F_\2}\wedge \hat F_\2\,,\cr
\hat H_\3&=&d\hat B_\2+\ft{1}{2}\hat F_\2\wedge\hat A_\1\,,\qquad \hat F_\2 = d\hat A_\1\,.\label{heterotic}
\eea
The low-energy effective Lagrangian of the noncritical bosonic string with a conformal anomaly term is
\bea
\check{\mathcal{L}}_{10}&=&\big(\check{R}-8g^2e^{\frac{1}{2}\check{\phi}}\big)\check{*}\oneone-
\ft{1}{2}\check{*}d\check{\phi}\wedge d\check{\phi}
-\ft{1}{2}e^{-\check{\phi}}\check{*}\check{H}_\3\wedge \check{H}_\3\,,\qquad \check{H}_\3=d\check{B}_\2\,. \label{10D anomalous string theory}
\eea
These two apparently very different theories can be reduced to the same $D=9$ Lagrangian with an appropriate Kaluza-Klein reduction ansatz. For the bosonic string, we can perform the ordinary Kaluza-Klein $S^1$ reduction, but keeping the Kaluza-Klein and winding vectors the same. In doing so, we can truncate out consistently one scalar field. The reduction ansatz is
\bea
d\check{s}_{10}^2&=&e^{-\frac{\bar{\phi}}{14a_9}}d\bar{s}_9^2+e^{\frac{\bar{\phi}}{2a_9}}\big(du-\frac{1}{\sqrt{2}}\bar{A}_{(1)}\big)^2\,,\cr
\check{\phi}&=&-\frac{1}{a_9}\bar{\phi},\qquad \check{B}_{(2)}=\bar{B}_{(2)}+\frac{1}{\sqrt{2}}\bar{A}_{(1)}\wedge du\,.\label{bosonickk}
\eea
Here, the coordinate $u$ is the internal coordinate of which all ten-dimensional fields are independent. The reduction on the heterotic theory is less standard, making use of some subtle gauging of the constant shift symmetry of the heterotic theory. The technique was developed in \cite{Kerimo:2003am,Kerimo:2004md, Kerimo:2004qx}, which constructed supergravities with Minkowski$\times$sphere vacua. The reduction ansatz is the warped Kaluza-Klein $\mathbb R$ reduction \cite{Kerimo:2003am,Kerimo:2004md, Kerimo:2004qx,Ma:2025mvo}
\bea
d\hat{s}_{10}^2&=&e^{2mu}\Big(e^{-\frac{\bar{\phi}}{14a_9}}d\bar{s}_9^2+e^{\frac{\bar{\phi}}{2a_9}}du^2\Big),\qquad a_9=-\sqrt{\frac{8}{7}}\,,\cr
\hat{\phi}&=&-\frac{1}{a_9}\bar{\phi}-4mu,\qquad \hat{B}_{(2)}=\bar{B}_{(2)},\qquad \hat{A}_{(1)}=\bar{A}_{(1)}\,.
\eea
Note that the conformal factor $e^{2m u}$ breaks the $U(1)$ isometry along the $u$ direction and hence it is natural to turn off the Kaluza-Klein vector in the reduction ansatz. Under these respectively reductions, both the heterotic and bosonic string theories reduce to the same $D=9$ theory, with the Lagrangian
\bea
\bar{\mathcal{L}}_{9}&=&
\big(\bar{R} -64m^2e^{-\frac{1}{2}a_9\bar{\phi}}\big){\bar *\oneone}
-\ft12 {\bar *d\bar \phi}\wedge d\bar\phi -\ft12 e^{a_9\bar{\phi}}
{\bar * \bar H}_\3\wedge \bar H_\3
-\ft12 e^{\frac{1}{2}a_9\bar{\phi}} {\bar *\bar F}_\2\wedge \bar F_\2\,,\cr
\bar{H}_{\3}&=&d\bar{B}_{\2}+\ft{1}{2}\bar{F}_{\2}\wedge \bar{A}_{\1},\qquad \bar{F}_{\2}=d\bar{A}_{\1},\qquad g^2=8m^2\,. \label{9D SS model}
\eea
Since both the reductions are consistent, we can thus use this $D=9$ theory as a bridge to map the relevant solutions in the heterotic theory to those in the bosonic string and vice versa.

In superstring theories, bosonic background solutions that preserve a certain fraction of supersymmetry play a particularly important role. To preserve supersymmetry in the heterotic string, there must exist a Killing spinor $\epsilon$ that satisfies the following set of the Killing spinor equations \cite{Bergshoeff:1981um,Chapline:1982ww}:
\bea
\delta \hat \psi_M &=& \hat D_M \hat \epsilon - \fft1{96} e^{-\fft12\hat \phi}\big(
\Gamma_M \Gamma^{PQR} - 12 \delta_M^P \Gamma^{QR}\big) \hat H_{PQR}\, \hat \epsilon=0\,,\nn\\
\delta \hat \chi &=& -\fft12 \Gamma^M \partial_M\hat{\phi}\,\hat \epsilon -
\fft{1}{24} e^{-\fft12 \hat \phi} \Gamma^{MNP} \hat H_{MNP}\,\hat \epsilon=0\,,\nn\\
\delta \hat \lambda &=& -\fft12 e^{-\fft14 \hat\phi} \Gamma^{MN} \hat F_{MN}\, \hat \epsilon =0\,.\label{susy}
\eea
However, there is no supersymmetry in the noncritical bosonic string. Nevertheless, it was shown \cite{Ma:2025mvo} that the theory admits Killing spinor equations, whose integrability condition gives rise to the full set of equations of motion. This is analogous to Einstein gravity in general dimensions, which admits Killing spinor equation $\nabla_\mu \epsilon=0$, whose integrability condition leads to Einstein's equation of motion, even though the Einstein theory is not supersymmetric in general dimensions. 

The existence of Killing spinors allows one to construct pseudo-supergravity extension of the bosonic string, where the pseudo-supergravity Lagrangian is invariant under pseudo-supersymmetric transformation rules at the quadratic order of the fermion fields. (True supersymmetric Lagrangians would be invariant at the full quartic fermion orders.) The pseudo-supersymmetric transformation rules and the corresponding Killing spinor equations are \cite{Lu:2011zx,Liu:2012jra,Lu:2011ku}
\bea
\delta \check \psi_M &=& \check D_M \check \epsilon - \fft1{96} e^{-\fft12\check \phi}\big(
\Gamma_M \Gamma^{PQR} - 12 \delta_M^P \Gamma^{QR}\big) \check H_{PQR}\, \check \epsilon -\fft{\rm i}{4\sqrt2} g\, e^{\fft14 \check\phi} \Gamma_M \check \epsilon =0\,,\nn\\
\delta \check \chi &=& -\fft12 \Gamma^M \partial_M\check{\phi}\,\check \epsilon -
\fft{1}{24} e^{-\fft12 \check \phi} \Gamma^{MNP} \check H_{MNP}\,\check \epsilon - \fft{\rm i}{\sqrt2} g\, e^{\fft14 \check \phi}\check \epsilon=0\,.\label{susy2}
\eea
It should be pointed out that we must be able to find dual pairs by the virtual of our construction. What is nontrivial is that such a dual pair of solutions would both admit Killing spinors in their respective theories. The corresponding Killing spinors must take the following form
\be
\hat \epsilon \sim e^{\fft12 m u} \hat \epsilon_0\,,\qquad \longleftrightarrow\qquad 
\check \epsilon \sim \check \epsilon_0\,,\qquad \hbox{with}\qquad \partial_u \hat \epsilon_0=0=
\partial_u \check\epsilon_0\,.
\ee
The purpose of this paper is to construct examples of such BPS and pseudo-BPS dual pairs, since they suggest that the proposed duality may also be extended to the fermionic sector.

\subsection{BPS and pseudo-BPS vacua}

In this subsection, we review the vacua solutions constructed in \cite{Ma:2025mvo} and add some new discussions on the Killing spinors. First, the noncritical bosonic string admits a Mink$_4\times S^3\times S^3$ vacuum:
\bea
d\check{s}_{10}^2&=&-dt^2 + dx^2 + dy^2 + dz^2+\frac{1}{g^2}\left(d\Omega_3^2+d\widetilde{\Omega}_3^2\right), \cr
\check{H}_{\3}&=&\frac{2}{g^2}\left(\omega_{\3}+\widetilde{\omega}_{\3}\right),\qquad \check\phi=0\,.
\eea
A key observation is that the direct product $S^3\times S^3$ can be viewed as a $U(1)$ bundle over $T^{1,1}$, so that the solution can be expressed as
\bea
d\check{s}_{10}^2 &=& \eta_{\mu\nu}dx^\mu dx^\nu+\frac{1}{8g^2}ds^2_{T^{1,1}}+\frac{1}{8g^2}\left(d\chi_2+\cos\theta_1 d\varphi_1-\cos\theta_2 d\varphi_2\right)^2,\qquad \check \phi =0\,.\cr
ds^2_{T^{1,1}} &=& \big(d\chi_1+\cos\theta_1 d\varphi_1+\cos\theta_2 d\varphi_2\big)^2
+2(d\theta_1^2 + \sin^2\theta_1\, d\varphi_1^2 + d\theta_2^2 + \sin^2\theta_2\, d\varphi_2^2)\,,\cr
\check{H}_{\3}
&=&\frac{1}{8g^2}\left(\sin\theta_1\,d\theta_1 \wedge d\varphi_1 -\sin\theta_2\,d\theta_2 \wedge d\varphi_2 \right)\wedge d\chi_2\nn\\
&&+\frac{1}{8g^2}\left(\sin\theta_1\,d\theta_1 \wedge d\varphi_1 +\sin\theta_2\,d\theta_2 \wedge d\varphi_2 \right)\wedge d\chi_1\,.\label{bvacsol2}
\eea
This expression of the solution fits the reduction ansatz of the bosonic/heterotic duality discussed in the previous subsection, with $\chi_2\sim u$.  Applying the duality, one obtains the dual solution in the heterotic theory \cite{Ma:2025mvo}
\bea
d\hat s_{\rm str}^2 &=& e^{\fft12\hat\phi} d\hat s_{\rm Ein}^2=\eta_{\mu\nu}dx^\mu dx^\nu+\frac{1}{64m^2}ds^2_{T^{1,1}}+du^2\,,\nn\\
d\hat{s}_{\rm Ein}^2&=&e^{2m u}\Big(\eta_{\mu\nu}dx^\mu dx^\nu+\frac{1}{64m^2}ds^2_{T^{1,1}}+du^2
\Big),\qquad \hat{\phi}=-4m u\,,\label{hetsol}\\
\hat{B}_{\2}&=&-\frac{1}{64m^2}\big(\cos\theta_1 d\varphi_1+\cos \theta_2 d\varphi_2\big)\wedge d\chi_1\,,\quad \hat{A}_{\1}=-\frac{\sqrt{2}}{8m}\big(\cos\theta_1 d\varphi_1-\cos\theta_2 d\varphi_2\big).\nn
\eea
In the string frame, the metric of the heterotic solution is a direct product of Mink$_4\times \mathbb R\times T^{1,1}$. In the Einstein frame, the metric becomes a warped product, with the warp factor depending on the $\mathbb R$ coordinate $u$.

It turns out that the new heterotic solution \eqref{hetsol} is actually supersymmetric, preserving $1/4$ of the supersymmetry. Specifically, for the vielbein choice given in \cite{Ma:2025mvo}, Killing spinors are \cite{Ma:2025mvo}
\be
\hat \epsilon = e^{\fft12 m u}\,\Big(e^{\fft{\rm i}{2} \chi_1}\hat \epsilon_+ + e^{-\fft{\rm i}{2} \chi_1}\hat \epsilon_-\Big)\,,
\ee
where $\hat \epsilon_\pm$ are constant spinors satisfying the projection
\be
\Gamma^{45}\hat \epsilon_\pm = \Gamma^{67}\hat \epsilon_\pm=\Gamma^{89} \hat \epsilon_\pm
=\mp {\rm i}\, \hat \epsilon_\pm\,.\label{hetvacks}
\ee
Note that we present the Killing spinors by the appropriate $\Gamma$ matrices projections, instead of the explicit 32 spinor components so that our results are independent of the choice of $\Gamma$ matrices. (However, there is still a residue ambiguity of the sign choice of each $\Gamma$ matrix that can lead to an opposite sign of the eigenvalues in \eqref{hetvacks}.)

At first sight, it is very unusual that we can get a supersymmetric solution in the heterotic theory when its dual origin is the vacuum of the noncritical bosonic string. Arguments were given in \cite{Ma:2025mvo} that although the bosonic string theory is not supersymmetric, it is pseudo-supersymmetric in that the theory admits Killing spinor equations that are consistent with the full bosonic equations of motion. Based on the Killing spinor structures of $S^3$, it was argued that half of the Killing spinors in the bosonic string survive the duality transformation and the other half do not.  Since there was no explicit demonstration given in \cite{Ma:2025mvo}, we complete the discussion here. For the bosonic vacuum solution \eqref{bvacsol2}, we adopt the following natural choice of the vielbein
\bea
&& \check e^{\underline\mu}=dx^\mu\,,\quad \check e^4= \fft{1}{2g}\,d\theta_1\,,\quad
\check e^5=\fft{1}{2g} \sin\theta_1\,d\varphi_1\,,\quad \check e^6= \fft{1}{2g}\,d\theta_2\,,\quad
\check e^7=\fft{1}{2g} \sin\theta_2\,d\varphi_2\,,\label{d10bvacviel}\\
&&\check e^8=\fft{1}{2\sqrt2\,g} (d\chi_1 + \cos\theta_1\, d\varphi_1 + \cos\theta_2\, d\varphi_2)\,,\quad
\check e^9 = \fft{1}{2\sqrt2\,g} (d\chi_2 + \cos\theta_1\, d\varphi_1 - \cos\theta_2\, d\varphi_2)\,.\nn
\eea
The corresponding non-vanishing spin connection components are given by \eqref{bstrppsp}, but with $H_Q=1=H_K$. In this paper, we present all the relevant spin connections in the appendix.

In ten dimensions, $\epsilon$ has a total of 32 components. Substituting the bosonic background into the Killing spinor equations \eqref{susy2}, we find that there are a total of 16 Killing spinors, given by
\be
\check \epsilon = e^{\fft12{\rm i} \chi_1} \check \epsilon_{1,+} +e^{-\fft12{\rm i} \chi_1} \check \epsilon_{1,-}+  e^{\fft12{\rm i} \chi_2} \check \epsilon_{2,+} + e^{-\fft12{\rm i} \chi_2} \check \epsilon_{2,-} \,,
\ee
where $\check \epsilon_{i,\pm}$ are constant spinors satisfying the projections
\bea
\Gamma^{45}\, \check \epsilon_{1,\pm} = \Gamma^{67}\, \check \epsilon_{1,\pm} = -
{\rm i}\, \Gamma^8\, \check \epsilon_{1,\pm} &=& \mp {\rm i}\, \check \epsilon_{1,\pm}\,,\nn\\
\Gamma^{45}\, \check \epsilon_{2,\pm} = -\Gamma^{67}\, \check \epsilon_{2,\pm} = -
{\rm i}\, \Gamma^9\, \check \epsilon_{2,\pm} &=& \mp {\rm i}\, \check \epsilon_{2,\pm}\,.\label{bvacks}
\eea
The picture is now crystal clear. There are 8 Killing spinors that depend on $\chi_2$, corresponding to the $u$ coordinate in the duality map, and hence they could not survive the bosonic/heterotic duality map. The 8 Killing spinors that are independent of $\chi_2$ will survive the duality map and become Killing spinors of the heterotic theory. (Keep in mind that a further chirality projection reduces the Killing spinor numbers to four.)
Thus the heterotic vacuum is truly supersymmetric, preserving $1/4$ of the supersymmetry.

In this paper, we shall construct more examples of BPS and pseudo-BPS dual pairs in the respective heterotic and noncritical bosonic strings.

\section{The $S^3 \times S^3$ reduction and $D=4$ BPS solutions}

The vacuum dual pair discussed in the previous section suggests that it is easier to construct the solutions in the noncritical bosonic string. To obtain more possible solutions, we consider $S^3\times S^3$ reduction of the bosonic string. For simplicity, we shall keep only the singlets under the $S^3\times S^3$ transitive group action. (The nontrivial sphere reduction that yields $SU(2)\times SU(2)$ were obtained in \cite{Ma:2025mvo}.) This reduction can be generalized to noncritical bosonic strings in general dimensions, and therefore we give a general discussion.

\subsection{The reduction}

We consider the low-energy effective theory of the noncritical bosonic string in general $D$ dimensions. The Lagrangian is given by
\bea
\check{\mathcal{L}}_D=\sqrt{-\check g} \Big(\check{R}_D-\ft{1}{2}(\partial\check{\phi})^2-\ft{1}{12}e^{a_D\check{\phi}}\check{H}_{\3}^2
-8g^2e^{-\frac{1}{2}a_D\check{\phi}}\Big),\qquad a_D=-\sqrt{\frac{8}{D-2}}\,.\label{D theory}
\eea
The reduction ansatz that retains all the singlets of the $SO(4)\sim SU(2)\times SU(2)$ isometry is \cite{Bremer:1998zp}
\bea
d\check{s}_D^2 &=& e^{2\alpha\varphi}ds_{D-6}^2+e^{2\beta\varphi}ds_6^2\,,\qquad
ds_6^2=\frac{1}{g^2}\left(d\Omega_3^2+d\widetilde{\Omega}_3^2\right),\cr
\check{H}_{\3}&=&H_{\3}+\frac{2}{g^2}\left(\omega_{\3}+\tilde{\omega}_{\3}\right)\,,\qquad \beta=-\ft16(D-8)\alpha,\quad \alpha=\sqrt{\frac{3}{(D-2)(D-8)}}\,.
\label{S3 S3 reduction 1}
\eea
The lower $(D-6)$-dimensional reduced theory is
\bea
\mathcal{L}_{D-6}&=&R_{D-6}-\ft{1}{2}(\partial\check{\phi})^2-\ft{1}{2}(\partial\varphi)^2
-\ft{1}{12}e^{-4\alpha\varphi+a_D\check{\phi}}H_{\3}^2\cr
&&+12g^2e^{\frac{1}{3}(D-2)\alpha\varphi}-4g^2e^{(D-6)\alpha\varphi+a_D\check{\phi}}-
8g^2e^{2\alpha\varphi-\frac{1}{2}a_D\check{\phi}}\,.
\eea
It is advantageous to make an orthonormal transformation on the scalars and  define $\phi_{1,2}$ as
\be
-4\alpha\varphi+a_D\check{\phi}=-\sqrt{16\alpha^2+a_D^2}\phi_1,\quad  -a_D\varphi-4\alpha\check{\phi}=\sqrt{16\alpha^2+a_D^2}\phi_2
\ee
and the $(D-6)$-dimensional theory changes to
\bea
\mathcal{L}_{D-6}&=&\sqrt{-g} \Big(R_{D-6}-\ft{1}{2}(\partial\phi_1)^2-\ft{1}{2}(\partial\phi_2)^2-
\ft{1}{12}e^{-2\sqrt{\frac{2}{D-8}}\phi_1}H_{\3}^2\cr
&&-4g^2e^{\sqrt{\frac{2}{D-8}}\phi_1}\big(1-e^{-\frac{\phi_2}{\sqrt{3}}}
\big)^2\big(2+e^{-\frac{\phi_2}{\sqrt{3}}}\big)\Big).
\eea
The scalar potential has a fixed point at $\phi_2=0$, in which case, the scalar potential vanishes. This is consistent with the earlier observation that the vacuum should be Mink$_4$. Note that the scalar field $\phi_2$ is massive, satisfying the linearized equation
\be
(\Box -8 g^2) \phi_2=0\,.
\ee
Therefore, the excitation of the $\phi_2$ field will necessarily break the pseudo-supersymmetry and we set $\phi_2=0$ consistently.  In what follows, we will denote the remaining scalar field $\phi_1$ simply as $\phi$. Setting $D=10$, we obtain the four-dimensional Lagrangian
\be
\mathcal{L}_4=\sqrt{-g} \Big( R-\ft{1}{2}(\partial\phi)^2-\ft{1}{12}e^{-2\phi} H_{\3}^2\Big).\label{d4lag}
\ee
This Lagrangian is part of $D=4$, ${\cal N}=1$ supergravity obtained from the warped $\mathbb R\times T^{1,1}$ reduction from heterotic theory. The full theory includes also the $SU(2)\times SU(2)$ Yang-Mills fields, whose gauge symmetry has an origin of the isometry group of the $T^{1,1}$ internal space \cite{Ma:2025mvo}. Such higher dimensional geometric origin for Yang-Mills in flat space is rare in literature and the only known other example was given by \cite{Gibbons:2003gp}. The 3-form field strength is Hodge dual to an axion, which forms a complex scalar supermultiplet with the dilaton $\phi$. The Killing spinor equations associated with the supersymmetric transformation rules are
\be
\delta \chi =\ft12\nabla_a \phi \Gamma^a \epsilon+ \ft1{12} e^{-\phi} H_{abc} \Gamma^{abc} \epsilon=0\,,\qquad
\delta \psi_a = \nabla_a \epsilon + \ft1{48} e^{-\phi} H_{bcd} [\Gamma^{bcd},\Gamma_a]\epsilon=0\,.
\ee
For ${\cal N}=1$ chiral theory, a chiral projection is also understood.  In the context of pseudo-supergravity as the dimension reduction of $D=10$ noncritical bosonic string, these linear transformation rules can also be obtained directly from \eqref{susy2}. Owing to the consistency of all our dimensional reductions, the construction of $D=4$ BPS states becomes equivalent to the constructions of BPS or pseudo-BPS solutions in heterotic or noncritical bosonic strings.

It should be emphasized that the simplest way of getting the $D=4$ theory \eqref{d4lag} is to simply reduce the heterotic theory on torus or Calabi-Yau threefolds and consistently truncate out the extra fields. Our method on the other hand arrives at $D=4$ via very different routes, either from the bosonic string or the heterotic theory, with different internal spaces. The lifting of the $D=4$ solutions can thus give rise to richer structures of higher dimensional solutions.

\subsection{BPS cosmic strings}

The $D=4$ theory admits an electric cosmic string solution, where the antisymmetry tensor $B_\2$ carries electric charges. Including also the pp-wave component along the string direction, we find that the solution is given by
\bea
ds_{4}^2&=&-H_K^{-1}dt^2+ H_K\big[dx+ (H_K^{-1}-1)dt\big]^2
+H_Q\big(dy^2+dz^2 \big),\cr
B_\2&=&H_Q^{-1}dt\wedge dx\,,\qquad
\phi=-\log H_Q\,.
\eea
The spacetime is split into the worldsheet $(x,t)$ directions and the transverse $(y,z)$ directions.
Both \( H_Q \) and \( H_K \) are harmonic functions defined in the transverse \( y \)-\( z \) plane, {\it i.e.}
\be
\big(\partial_y^2+\partial_z^2\big)H_Q=0\,,\quad   \big(\partial_y^2+\partial_z^2\big)H_K=0\,.\label{HQ HK}
\ee
Turning off the pp-wave component by setting $H_K=1$, the cosmic string solution was extensively studied in  \cite{Greene:1989ya}. It was shown that $H_Q$ can be solved in terms of Dedekind's functions and regularity requires the superposition of 24 cosmic strings.

For calculating the Killing spinors, we adopt the vielbein
\be
e^0 = \fft{1}{\sqrt{H_K}} dt\,,\quad e^1 = \sqrt{H_K} (dx + (H_K^{-1} -1) dt)\,,\quad
e^2 = \sqrt{H_Q}\, dy\,,\quad e^3 = \sqrt{H_Q}\, dz\,.\label{d4viel1}
\ee
We find that the Killing spinors are given by
\be
\epsilon = H_K^{-\fft14} \epsilon_0\,,\qquad \Gamma^{01} \epsilon_0 = \epsilon_0\,.
\ee
In other words, the solution preserves $1/2$ of the supersymmetry. It is worth commenting here that the above $\Gamma^{01}$ projection is required by both the $H_Q$ and $H_K$ components. If either $H_Q=1$ or $H_K=1$, the solution breaks $1/2$ of the supersymmetry. When they are both turned on, they either preserving $1/2$ of the supersymmetry together, or they break all the supersymmetry, depending on the signs of the string charge or pp-wave direction.

\subsection{BPS instanton solution}

The magnetic dual to the electric cosmic solution in four dimensions is an instanton, where the spacetime requires to be Euclidean. We find that the Euclidean instanton solution is
\bea
ds_4^2&=&dr^2+ \ft14 r^2 (\sigma_1^2 + \sigma_2^2 + \sigma_3^2)\,,\qquad
H_\3 =2a\omega_\3 = \ft14a \sigma_1\wedge\sigma_2\wedge \sigma_3\,,\cr
\phi&=&\log H_a\,,\qquad H_a=1+\frac{a}{r^2}\,.\label{Euclidean instanton solution}
\eea
Here, we used the $SU(2)$-left invariant 1-form $\sigma_i$ to describes the unit $S^3$ metric, $d\Omega_3^2 = \fft14 (\sigma_1^2 + \sigma_2^2 + \sigma_3^2)$, where $d\sigma_i=-\fft12 \epsilon_{ijk} \sigma_j\wedge\sigma_k$.  These 1-forms can be expressed in terms of three Euler angles, given by
\be
\sigma_1=\cos\psi d\theta+\sin\psi\sin\theta d\varphi\,,\quad \sigma_2=-\sin\psi d\theta+\cos\psi\sin\theta d\varphi\,,\quad \sigma_3=d\psi+\cos\theta d\varphi\,.\label{su21form}
\ee
It is natural to adopt the vielbein choice
\be
e^0 = dr\,,\qquad e^i=\ft12r \sigma^i\,,
\ee
for which, the spin connections are $\omega_{0i} = -e^i/r$ and $\omega_{ij} =-\epsilon_{ijk} e^k/r$. In this vielbein choice, the Killing spinors are constant, subject to the chirality condition
\be
\Gamma^{0123} \epsilon =\epsilon\,.
\ee
Note that the bosonic solution also allows to have the negative branch, $H_\3=-2a\omega_\3$. In this case, we appear to have no Killing spinors. A careful examination indicates that the Killing spinors are more complicated in this case,  and they are no longer constants but depend on the Euler angles. Alternatively, we can make a different choice of the vielbein, namely $e^i=-\ft12 r\sigma_i$, then the Killing spinors are constants again. In either choice of the vielbein, the negative branch Killing spinors satisfy the anti-chiral projection $\Gamma^{0123} \epsilon =-\epsilon$. In other words, if the $D=4$ is chiral, then only one branch of the instanton solution is supersymmetric.

It is worth pointing out that in the string frame, $ds_{\rm str} = H_a^2 ds_4^2$, the metric describes a wormhole, with the inside ($r<0$) identical to the outside ($r>0$). This is analogous to the D-instanton of type IIB supergravity \cite{Bergshoeff:1998ry}. The wormhole throat at $r=0$ is in the strong coupling region, where $g_{\rm str} =e^{\phi}\rightarrow \infty$.

\section{Lifting back to $D=10$}

In the previous section, we obtained both BPS string and instanton solutions in four dimensions. Since the theory can be obtained from consistent reductions of either heterotic or bosonic string theories, both solutions have ten-dimensional origins.

\subsection{Pseudo-BPS states in the noncritical string theory}

The cosmic string with the pp-wave, after lifting to the $D=10$, becomes a solution in the noncritical bosonic string theory. It is given by
\bea
d\check{s}_{10}^2&=&H_Q^{-3/4}\Big\{-H_K^{-1}dt^2+H_K\big[dx+(H_K^{-1}-1)dt\big]^2\Big\}
+H_Q^{1/4}\Big[dy^2+dz^2+\frac{1}{g^2}\big(d\Omega_3^2+d\widetilde{\Omega}_3^2\big)
\Big],\cr
\check{H}_\3&=&dH_Q^{-1}\wedge dt\wedge dx + \fft{2}{g^2}(\omega_\3 + \widetilde\omega_\3)\,,\qquad
\check{\phi} =-\frac{1}{2}\log H_Q\,.\label{d10bstring}
\eea
Note that the eight-dimensional transverse space is split into a direct product of Euclidean 2-plane and $S^3\times S^3$. For the most general solution, $H_Q, H_K$ are harmonic functions in the whole transverse space. For simplicity, we shall let \( H_Q \) and \( H_K \) be the harmonic functions defined on the \( y \)-\( z \) plane, as in \eqref{HQ HK}.

To show the solution admits Killing spinors, we can simply choose a vielbein base associated with the coordinate base above. However, to illustrate explicit how many Killing spinors can survive the bosonic/heterotic string duality, it is advantageous to write the $S^3\times S^3$ as a $U(1)$ bundle over $T^{1,1}$, as in the case for the Mink$_4\times S^3\times S^3$. In this new coordinate system, the solution becomes
\bea
d\check{s}_{10}^2&=&H_Q^{-\fft34}\Big\{-H_K^{-1}dt^2+H_K\big[dx+(H_K^{-1}-1)dt\big]^2\Big\}\nn\\
&&+H_Q^{\fft14}\Big[dy^2+dz^2+\frac{1}{8g^2}ds^2_{T^{1,1}}+\frac{1}{8g^2}\left(d\chi_2+\cos\theta_1 d\varphi_1-\cos\theta_2 d\varphi_2\right)^2
\Big],\cr
\check{H}_\3&=&dH_Q^{-1}\wedge dt\wedge dx + \frac{1}{8g^2}\left(\sin\theta_1\,d\theta_1 \wedge d\varphi_1 -\sin\theta_2\,d\theta_2 \wedge d\varphi_2 \right)\wedge d\chi_2\cr
&&+\frac{1}{8g^2}\left(\sin\theta_1\,d\theta_1 \wedge d\varphi_1 +\sin\theta_2\,d\theta_2 \wedge d\varphi_2 \right)\wedge d\chi_1\,,\qquad
\check{\phi} =-\frac{1}{2}\log H_Q\,.\label{d10bstring2}
\eea
We now choose the vielbein base
\bea
&&\check e^{0}=H_Q^{-\fft38} H_K^{-\fft12} dt\,,\quad \check e^1 = H_Q^{-\fft38} H_K^{\fft12} (dx + (H_K^{-1} -1) dt)\,,\quad
\check e^2 = H_Q^{\fft18} dy\,,\quad \check e^3 = H_Q^{\fft18} dz\,,\cr
&&
\check e^4= \fft{H_Q^{\fft18}}{2g}\,d\theta_1\,,\quad
\check e^5=\fft{H_Q^{\fft18}}{2g} \sin\theta_1\,d\varphi_1\,,\quad \check e^6= \fft{H_Q^{\fft18}}{2g}\,d\theta_2\,,\quad
\check e^7=\fft{H_Q^{\fft18}}{2g} \sin\theta_2\,d\varphi_2\,,\label{d10bstrviel1}\\
&&\check e^8=\fft{H_Q^{\fft18}}{2\sqrt2\,g} (d\chi_1 + \cos\theta_1\, d\varphi_1 + \cos\theta_2\, d\varphi_2)\,,\quad
\check e^9 = \fft{H_Q^{\fft18}}{2\sqrt2\,g} (d\chi_2 + \cos\theta_1\, d\varphi_1 - \cos\theta_2\, d\varphi_2)\,.\nn
\eea
We find that the Killing spinors are given by
\be
\check \epsilon =H_Q^{-\fft3{16}} H_K^{-\fft14} \Big( e^{\fft12{\rm i} \chi_1} \check \epsilon_{1,+} +e^{-\fft12{\rm i} \chi_1} \check \epsilon_{1,-}+  e^{\fft12{\rm i} \chi_2} \check \epsilon_{2,+} + e^{-\fft12{\rm i} \chi_2} \check \epsilon_{2,-} \Big)\,,
\ee
where the constant spinors $\check \epsilon_{i,\pm}$ satisfy the projections \eqref{bvacks}, in addition to following extra condition
\be
\Gamma^{01} \check \epsilon_{i,\pm} =- \check \epsilon_{i,\pm}\,,\qquad i=1,2\,.
\ee
Thus, there are a total of 8 Killing spinors, preserving half of the total pseudo-supersymmetry. (A chiral projection can reduce the total number of Killing spinors by half.) There are four Killing spinors depending on the $\chi_1$ coordinates, and four Killing spinors depending on the $\chi_2$ coordinates.

The four-dimensional Euclidean instanton solution \eqref{Euclidean instanton solution} can also be lifted to become a solution of the noncritical bosonic string theory. It becomes
\bea
d\check s_{10}^2 &=& H_a^{\fft34} (dr^2 + \ft14 r^2 (\sigma_1^2 + \sigma_2^2 + \sigma_3^2)) +
\fft{H_a^{-\fft14}}{g^2} (d\Omega_3^2 + d\widetilde\Omega_3^2)\,,\cr
\check H_\3 &=& \ft14a \sigma_1\wedge\sigma_2\wedge \sigma_3 + \fft{2}{g^2}(\omega_\3 + \widetilde\omega_\3)\,,\qquad
\check \phi = \ft12 \log H_a\,,\qquad H_a= 1 + \fft{a}{r^2}\,.
\eea
As in the previous cases, we shall regard the $S^3\times S^3$ as a $U(1)$ bundle over $T^{1,1}$ and write the vielbein accordingly. The vielbein are
\bea
&&\check e^0 = H_a^{\fft38} dr\,,\qquad \check e^i = \ft12 r H_a^{\fft38} \sigma_i\,,\qquad \check e^4= \fft{H_a^{-\fft18}}{2g}\,d\theta_1\,,\cr
&&\check e^5=\fft{H_a^{-\fft18}}{2g} \sin\theta_1\,d\varphi_1\,,\qquad \check e^6= \fft{H_a^{-\fft18}}{2g}\,d\theta_2\,,\qquad
\check e^7=\fft{H_a^{-\fft18}}{2g} \sin\theta_2\,d\varphi_2\,,\label{d10bstrviel2}\\
&&\check e^8=\fft{H_a^{-\fft18}}{2\sqrt2\,g} (d\chi_1 + \cos\theta_1\, d\varphi_1 + \cos\theta_2\, d\varphi_2)\,,\quad
\check e^9 = \fft{H_a^{-\fft18}}{2\sqrt2\,g} (d\chi_2 + \cos\theta_1\, d\varphi_1 - \cos\theta_2\, d\varphi_2)\,.\nn
\eea
We find that there are 8 Killing spinors, given by
\be
\check \epsilon =H_a^{-\fft1{16}} \Big( e^{\fft12{\rm i} \chi_1} \check \epsilon_{1,+} +e^{-\fft12{\rm i} \chi_1} \check \epsilon_{1,-}+  e^{\fft12{\rm i} \chi_2} \check \epsilon_{2,+} + e^{-\fft12{\rm i} \chi_2} \check \epsilon_{2,-} \Big)\,,
\ee
where the constant spinors $\check \epsilon_{i,\pm}$ satisfy the projections \eqref{bvacks}, in addition to a further projection due to the instanton charge
\be
\Gamma^{0123} \check \epsilon_{i,\pm} = \check\epsilon_{i,\pm}\,,\qquad i=1,2.
\ee
Thus we see that all of our new solutions in the bosonic string theory admit Killing spinors, with half depending on $\chi_1$ and half depending on $\chi_2$. As we have discussed, their duals in the heterotic string theory then must also admit Killing spinors. In particular, the Killing spinors independent of $\chi_2$ will survive the duality map. We shall examine this in the next subsection.

\subsection{New BPS states in the heterotic theory}

The four-dimensional BPS solutions can also be lifted back to become solutions in the heterotic theory, either via the bosonic/heterotic duality we discussed in Section 2, or by the reduction ansatz given in \cite{Ma:2025mvo}, but with the $SU(2)\times SU(2)$ turned off.

The cosmic string with pp-wave becomes a warped solution
\bea
d\hat{s}_{10}^2&=&e^{2mu}\Bigg\{
H_Q^{-3/4}\Big\{-H_K^{-1}dt^2+H_K\big[dx+(H_K^{-1}-1)dt\big]^2\Big\}\cr
&&+H_Q^{1/4}\Big[dy^2+dz^2+\frac{1}{64m^2}ds^2_{T^{1,1}}+du^2
\Big]
\Bigg\}\,, \cr
\hat{B}_{\2}&=&H_Q^{-1}dt\wedge dx-\frac{1}{64m^2}\big(\cos\theta_1 d\varphi_1+\cos\theta_2 d\varphi_2 \big)\wedge d\chi_1\,,\cr
\hat{A}_\1&=&-\frac{1}{4\sqrt{2}m}\big(\cos\theta_1 d\varphi_1-\cos\theta_2 d\varphi_2\big)\,,\qquad \hat{\phi}=-\frac{1}{2}\log H_Q-4mu\,.\label{pp wave heterotic}
\eea
We see that the transverse space continues to be eight-dimensional, comprising a $(y,z)$ 2-plane, $T^{1,1}$ and a warped $u$ direction. This solution preserves $1/8$ of supersymmetry. To verify this, we adopt the following vielbein choice:
\bea
\hat{e}^0&=&e^{mu}H_Q^{-\fft38}H_K^{-\fft12}dt\,,\qquad \hat{e}^1=e^{mu}H_Q^{-\fft38}H_K^{\fft12}\big[dx+(H_K^{-1}-1)dt\big]\,,\cr
\hat{e}^2&=&e^{mu}H_Q^{\fft18}dy\,,\qquad \hat{e}^3=e^{mu}H_Q^{\fft18}dz\,,\qquad \hat{e}^4=\frac{\sqrt{2}e^{mu}}{8m}H_Q^{\fft18}d\theta_1\,,\cr  \hat{e}^5 &=&\frac{\sqrt{2}e^{mu}}{8m}H_Q^{\fft18}\sin\theta_1 d\varphi_1\,,\quad
\hat{e}^6 =\frac{\sqrt{2}e^{mu}}{8m}H_Q^{\fft18}d\theta_2\,,\quad
\hat{e}^7=\frac{\sqrt{2}e^{mu}}{8m}H_Q^{\fft18}\sin\theta_2 d\varphi_2\,,\cr \hat{e}^8&=&\frac{e^{mu}}{8m}H_Q^{\fft18}\big(d\chi_1+\cos\theta_1 d\varphi_1+\cos\theta_2 d\varphi_2\big),\qquad \hat{e}^9=e^{mu}H_Q^{\fft18}du\,.\label{d10hstrviel1}
\eea
It follows from the Killing spinor equations \eqref{susy} that there are a total of four Killing spinors, given by
\bea
\hat{\epsilon}=e^{\frac{mu}{2}} H_K^{-\frac{1}{4}}H_Q^{-\frac{3}{16}}\Big(
e^{\frac{\mathrm{i}\chi_1}{2}}\hat{\epsilon}_++e^{-\frac{\mathrm{i}\chi_1}{2}}\hat{\epsilon}_-
\Big).
\eea
Here the constant spinors $\hat{\epsilon}_\pm$ satisfy the projections \eqref{hetvacks}. In addition, they also satisfy a further projection owing to the turning on of the string or pp-wave component:
\be
\Gamma^{01}\hat{\epsilon}_\pm=-\hat{\epsilon}_\pm,
\ee
In the heterotic string theory, it is understood that a further chiral projection should be imposed, but this does not affect the fact that the solution preserves $1/8$ of the total symmetry of the theory.

Analogously, the heterotic instanton solution is given by
\bea
d\hat{s}_{10}^2&=&e^{2mu}\Bigg\{
H_a^{\fft34}\big (dr^2+\ft14 r^2 (\sigma_1^2 + \sigma_2^2 + \sigma_3^2)\big)+H_a^{-\fft14}\Big[\frac{1}{64m^2}ds^2_{T^{1,1}}+du^2
\Big]\Bigg\}\,, \cr
\hat{H}_{\3}&=& \ft14a\,\sigma_1\wedge\sigma_2\wedge\sigma_3 +\frac{1}{64m^2}\big(\sin\theta_1\, d\theta_1\wedge d\varphi_1+\sin\theta_2\, d\theta_2\wedge d\varphi_2 \big)\wedge d\chi_1\cr
&& + \fft{1}{64m^2} \Big(\sin\theta_1 \cos\theta_2\, d\theta_1 -
\sin\theta_2 \cos\theta_1\, d\theta_2\Big) \wedge d\varphi_1 \wedge d\varphi_2\,,\cr
\hat{F}_\2&=&\frac{1}{4\sqrt{2}m}\big(\sin\theta_1\, d\theta_1\wedge d\varphi_1-\sin\theta_2\, d\theta_2\wedge  d\varphi_2\big)\,,\qquad \hat{\phi}=\frac{1}{2}\log H_a-4mu\,.\label{Euclidean instanton heterotic}
\eea
We choose the following vielbein to describe the Euclidean instanton solutions
\bea
\hat{e}^0&=&e^{mu}H_a^{\fft38}dr\,,\quad \hat{e}^1=\frac{r}{2}e^{mu}H_a^{\fft38}\sigma_1\,,\quad
\hat{e}^2=\frac{r}{2}e^{mu}H_a^{\fft38}\sigma_2\,,\quad \hat{e}^3=\frac{r}{2}e^{mu}H_a^{\fft38}\sigma_3, \cr
\hat{e}^4&=&\frac{\sqrt{2}e^{mu}}{8m}H_a^{-\fft18}d\theta_1\,,\qquad \hat{e}^5=\frac{\sqrt{2}e^{mu}}{8m}H_a^{-\fft18}\sin\theta_1 d\varphi_1,\qquad \hat{e}^6=\frac{\sqrt{2}e^{mu}}{8m}H_a^{-\fft18}d\theta_2\,,\cr
\hat{e}^7&=&\frac{\sqrt{2}e^{mu}}{8m}H_a^{-\fft18}\sin\theta_2 d\varphi_2\,,\qquad \hat{e}^8=\frac{e^{mu}}{8m}H_a^{-\fft18}\big(d\chi_1+\cos\theta_1 d\varphi_1+\cos\theta_2 d\varphi_2\big),\cr
\hat{e}^9&=&e^{mu}H_a^{-\fft18}du\,.\label{d10hstrviel2}
\eea
Again, the solution is supersymmetric, with the Killing spinors
\be
\hat{\epsilon}=e^{\frac{mu}{2}} {H_a^{-\frac{1}{16}}}\Big(e^{\frac{\mathrm{i}\chi_1}{2}}\hat{\epsilon}_{0,+}+
e^{-\frac{\mathrm{i}\chi_1}{2}}\hat{\epsilon}_{0,-}
\Big)\,,
\ee
where the constant spinors $\hat \epsilon_{0,\pm}$ satisfy the projections \eqref{hetvacks}, and in addition, they satisfy a further independent projection
\be
\Gamma^{0123} \hat\epsilon_{0,\pm} = \hat{\epsilon}_{0,\pm}\,.
\ee
It follows from \eqref{hetvacks} and the above, the $\Gamma^{11}$ projection of the Killing spinors is
\be
\Gamma^{11} \hat\epsilon_{0,\pm} =\pm {\rm i}\, \hat\epsilon_{0,\pm}\,.
\ee
Thus a specific choice of chirality, which we must impose in the heterotic theory,  further projects out half the Killing spinors, either $\hat \epsilon_{0,+}$ or $\hat \epsilon_{0,-}$, depending on the convention.

\section{Conclusions}

In this paper, we constructed two classes of new BPS solutions in heterotic supergravity. One describes a cosmic string with pp-wave and the other is a Euclidean instanton. Both solutions have $\mathbb R\times T^{1,1}$ transverse directions in the string frame. We demonstrate that these solutions are supersymmetric by constructing explicitly the Killing spinors.

We obtained these new heterotic BPS solutions by taking advantage of a recently proposed bosonic/heterotic duality and mapped the dual solutions in the noncritical bosonic string. These dual states in the bosonic string theory are pseudo-supersymmetric in that they also admit Killing spinors, solutions of the consistent Killing spinor equations of the bosonic string theory.

We should be cautious not to claim that the heterotic theory is equivalent to the bosonic string theory under the duality, since at the current form, the duality map requires the heterotic dilaton to have a specific linear dependence on an internal $\mathbb R$ coordinate. However, the map suggests that there may exist a consistent subsector of the heterotic theory that is related to the bosonic string. Our successful construction of the BPS and pseudo-BPS pairs lends support to this claim and the duality deserves further investigation.

\section*{Acknowledgement}

L.M.~is supported in part by National Natural Science Foundation of China (NSFC) grant No.~12447138, Postdoctoral Fellowship Program of CPSF Grant No.~GZC20241211 and the China Postdoctoral Science Foundation under Grant No.~2024M762338. H.L.~is supported in part by the National Natural Science Foundation of China (NSFC) grants No.~12375052 and No.~11935009.

\section*{Appendix}

\appendix

\section{Spin connections}

The vielbein 1-forms $e^a=e^{a}_\mu dx^\mu$ and the corresponding spin connection 1-forms $\omega^{a}_{{\color{white}2}b}=\omega^{a}_{\mu b} dx^\mu$ are necessarily in the Killing spinor calculations since the covariant derivative of a spinor $\epsilon$ is given by
\be
\nabla_\mu \epsilon = \partial_\mu \epsilon+ \ft14 \omega_{\mu ab} \Gamma^{ab} \epsilon\,.
\ee
In this appendix, we give the explicit non-vanishing components $\omega_{ab}$ ($a<b$) of the spin connections associated with various vielbeins in the main text. The remaining components are either zero or can be obtained from the antisymmetry property $\omega_{ba}=-\omega_{ab}$.

\subsection{4D Bosonic pp-wave}
For the vielbein \eqref{d4viel1}, the spin connections are
\bea
\omega_{01}&=&-\frac{1}{2H_K\sqrt{H_Q}}\big(H_K^{(1,0)}e^2+ H_K^{(0,1)}e^3\big)\,,\qquad \omega_{02}=\frac{ H_K^{(1,0)}}{2 H_K\sqrt{H_Q}}\big(e^0-e^1\big)\,,\cr
\omega_{03}&=&\frac{ H_K^{(0,1)}}{2 H_K\sqrt{H_Q}}\big(e^0-e^1\big)\,,\qquad \omega_{12}=-\frac{ H_K^{(1,0)}}{2 H_K\sqrt{H_Q}}\big(e^0-e^1\big)\,,\cr
\omega_{13}&&=-\frac{ H_K^{(0,1)}}{2 H_K\sqrt{H_Q}}\big(e^0-e^1\big)\,,\qquad
\omega_{23}=\frac{1}{2H_Q^{3/2}}\big(H_Q^{(0,1)}e^2- H_Q^{(1,0)}e^3\big)\,.
\eea
In this paper, $X^{(1,0)}$ and $X^{(0,1)}$ denote a derivative of $y$ and $z$ respectively. The form fields in the vielbein base are
\bea
H_{(3)}&=&-H_Q^{-\fft52} \big(H_Q^{(1,0)}e^2+H_Q^{(0,1)} e^3\big)\wedge e^0\wedge e^1\,,\cr
d\phi&=&-H_Q^{-\fft32}\big(H_Q^{(1,0)}e^2+H_Q^{(0,1)}e^3\big).
\eea

\subsection{Bosonic string with pp-wave}

For the vielbein \eqref{d10bstrviel1}, the spin connections are
\bea
\check{\omega}_{01}&=&-\frac{1}{2 H_KH_Q^{1/8}}\big(H_K^{(1,0)}\check{e}^2+ H_K^{(0,1)}\check{e}^3\big),\quad \check{\omega}_{0 2} = -\frac{ H_K^{(1,0)}}{2 H_KH_Q^{1/8}}\Big[\check{e}^1-\check{e}^0\big(1+\frac{3 H_K H_Q^{(1,0)}}{4 H_K^{(1,0)} H_Q}\big)\Big],\cr
\check{\omega}_{0 3}&=&-\frac{ H_K^{(0,1)}}{2 H_KH_Q^{1/8}}\Big[\check{e}^1-\check{e}^0\big(1+\frac{3 H_K H_Q^{(0,1)}}{4 H_K^{(0,1)} H_Q}\big)\Big],\cr
\check{\omega}_{1 2} &=&-\frac{ H_K^{(1,0)}}{2 H_KH_Q^{1/8}}\Big[\check{e}^0-\check{e}^1\big(1-\frac{3 H_K H_Q^{(1,0)}}{4 H_K^{(1,0)} H_Q}\big)\Big],\cr
\check{\omega}_{1 3}&=&-\frac{ H_K^{(0,1)}}{2 H_KH_Q^{1/8}}\Big[\check{e}^0-\check{e}^1\big(1-\frac{3 H_K H_Q^{(0,1)}}{4 H_K^{(0,1)} H_Q}\big)\Big],\quad
\check{\omega}_{2 3}=\frac{ 1}{8 H_Q^{9/8}}\big(H_Q^{(0,1)}\check{e}^2- H_Q^{(1,0)}\check{e}^3\big),\cr
\check{\omega}_{2 4}&=&-\frac{ H_Q^{(1,0)}}{8 H_Q^{9/8}}\check{e}^4,\quad
\check{\omega}_{2 5}=-\frac{ H_Q^{(1,0)}}{8 H_Q^{9/8}}\check{e}^5,\qquad \check{\omega}_{2 6}=-\frac{ H_Q^{(1,0)}}{8 H_Q^{9/8}}\check{e}^6,\quad
\check{\omega}_{2 7}=-\frac{ H_Q^{(1,0)}}{8 H_Q^{9/8}}\check{e}^7,\cr
\check{\omega}_{2 8}&=&-\frac{ H_Q^{(1,0)}}{8 H_Q^{9/8}}\check{e}^8,\quad
\check{\omega}_{2 9}=-\frac{ H_Q^{(1,0)}}{8 H_Q^{9/8}}\check{e}^9,\quad \check{\omega}_{3 4}=-\frac{ H_Q^{(0,1)}}{8 H_Q^{9/8}}\check{e}^4,\quad
\check{\omega}_{3 5}=-\frac{ H_Q^{(0,1)}}{8 H_Q^{9/8}}\check{e}^5,\cr
\check{\omega}_{3 6}&=&-\frac{ H_Q^{(0,1)}}{8 H_Q^{9/8}}\check{e}^6,\quad \check{\omega}_{3 7}=-\frac{ H_Q^{(0,1)}}{8 H_Q^{9/8}}\check{e}^7,\quad
 \check{\omega}_{3 8}=-\frac{ H_Q^{(0,1)}}{8 H_Q^{9/8}}\check{e}^8,\quad
\check{\omega}_{3 9}=-\frac{ H_Q^{(0,1)}}{8 H_Q^{9/8}}\check{e}^9,\cr
\check{\omega}_{4 5}&=&\frac{g }{ \sqrt{2} H_Q^{1/8}}\big(\check{e}^8+\check{e}^9-2\sqrt{2}\cot\theta_1\check{e}^5\big),\quad
\check{\omega}_{4 8}=\frac{g}{\sqrt{2} H_Q^{1/8}}\check{e}^5,\quad
\check{\omega}_{4 9}=\frac{g}{\sqrt{2} H_Q^{1/8}}\check{e}^5,\cr
\check{\omega}_{5 8}&=&-\frac{g}{\sqrt{2} H_Q^{1/8}}\check{e}^4,\quad \check{\omega}_{5 9}=-\frac{g}{\sqrt{2} H_Q^{1/8}}\check{e}^4,\quad
\check{\omega}_{6 7}=\frac{g }{ \sqrt{2} H_Q^{1/8}}\big(\check{e}^8-\check{e}^9-2\sqrt{2}\cot\theta_2\check{e}^7\big),\cr
\check{\omega}_{6 8}&=&\frac{g}{\sqrt{2} H_Q^{1/8}}\check{e}^7,\quad \check{\omega}_{6 9}=-\frac{g}{\sqrt{2} H_Q^{1/8}}\check{e}^7,\quad
\check{\omega}_{7 8}=-\frac{g}{\sqrt{2} H_Q^{1/8}}\check{e}^6,\quad
\check{\omega}_{7 9}=\frac{g}{\sqrt{2} H_Q^{1/8}}\check{e}^6.\label{bstrppsp}
\eea
The form fields are
\bea
\check{H}_{(3)}&=&-\frac{1}{H_Q^{11/8}} \big(H_Q^{(1,0)}\check{e}^2+H_Q^{(0,1)} \check{e}^3\big)\wedge\check{e}^0\wedge \check{e}^1\cr
&&+\frac{\sqrt{2}g}{H_Q^{3/8}}\big(\check{e}^4\wedge \check{e}^5+\check{e}^6\wedge \check{e}^7\big)\wedge \check{e}^8+\frac{\sqrt{2}g}{H_Q^{3/8}}\big(\check{e}^4\wedge \check{e}^5-\check{e}^6\wedge \check{e}^7\big)\wedge \check{e}^9,\cr
d\check{\phi}&=&-\frac{1}{2H_Q^{9/8}}\big(H_Q^{(1,0)}\check{e}^2+H_Q^{(0,1)}\check{e}^3\big)
\eea

\subsection{Bosonic Euclidean instanton solutions}

For the vielbein \eqref{d10bstrviel2}, the spin connections are
\bea
\check{\omega}_{0 1}&=&-\frac{8 H_a+3 r H_a'}{8 r H_a^{11/8}}\check{e}^1,\qquad \check{\omega}_{0 2}=-\frac{8 H_a+3 r H_a'}{8 r H_a^{11/8}}\check{e}^2,\qquad
\check{\omega}_{0 3}=-\frac{8 H_a+3 r H_a'}{8 r H_a^{11/8}}\check{e}^3,\cr
\check{\omega}_{0 4}&=&\frac{ H_a'}{8 H_a^{11/8}}\check{e}^4,\qquad
\check{\omega}_{0 5}=\frac{ H_a'}{8 H_a^{11/8}}\check{e}^5,\qquad
\check{\omega}_{0 6}=\frac{ H_a'}{8 H_a^{11/8}}\check{e}^6,\qquad
\check{\omega}_{0 7}=\frac{ H_a'}{8 H_a^{11/8}}\check{e}^7,\cr
\check{\omega}_{0 8}&=&\frac{ H_a'}{8 H_a^{11/8}}\check{e}^8\,,\qquad
\check{\omega}_{0 9}=\frac{ H_a'}{8 H_a^{11/8}}\check{e}^9,\qquad \check{\omega}_{1 2}=-\frac{1}{r H_a^{3/8}}\check{e}^3,\qquad
\check{\omega}_{1 3}=\frac{1}{r H_a^{3/8}}\check{e}^2,\cr
\check{\omega}_{2 3}&=&-\frac{1}{r H_a^{3/8}}\check{e}^1,\qquad
\check{\omega}_{4 5}=\frac{g}{\sqrt{2}}H_a^{1/8}\big(\check{e}^8+\check{e}^9-2\sqrt{2}\cot\theta_1\check{e}^5\big),\qquad \check{\omega}_{4 8}=\frac{g}{\sqrt{2}}H_a^{1/8}\check{e}^5,\cr
\check{\omega}_{4 9}&=&\frac{g}{\sqrt{2}}H_a^{1/8}\check{e}^5\,,\qquad
\check{\omega}_{5 8}=-\frac{g}{\sqrt{2}}H_a^{1/8}\check{e}^4,\qquad \check{\omega}_{5 9}=-\frac{g}{\sqrt{2}}H_a^{1/8}\check{e}^4, \cr
\check{\omega}_{6 7}&=&\frac{g}{\sqrt{2}}H_a^{1/8}\big(\check{e}^8-\check{e}^9-2\sqrt{2}\cot\theta_2\check{e}^7\big),\quad
\check{\omega}_{6 8}=\frac{g}{\sqrt{2}}H_a^{1/8}\check{e}^7,\qquad \check{\omega}_{6 9}=-\frac{g}{\sqrt{2}}H_a^{1/8}\check{e}^7,\cr
\check{\omega}_{7 8}&=&-\frac{g}{\sqrt{2}}H_a^{1/8}\check{e}^6,\qquad \check{\omega}_{7 9}=\frac{g}{\sqrt{2}}H_a^{1/8}\check{e}^6\,.
\eea
The form fields are
\bea
\check{H}_{(3)}&=&-\frac{H_a'}{H_a^{9/8}}\check{e}^1\wedge \check{e}^2\wedge \check{e}^3+\sqrt{2}gH_a^{3/8}\big(\check{e}^4\wedge \check{e}^5+\check{e}^6\wedge \check{e}^7\big)\wedge \check{e}^8\cr
&&+\sqrt{2}gH_a^{3/8}\big(\check{e}^4\wedge \check{e}^5-\check{e}^6\wedge \check{e}^7\big)\wedge \check{e}^9,\cr
d\check{\phi}&=&\frac{H_a'}{2H_a^{11/8}}\check{e}^0.
\eea

\subsection{Heterotic string with pp-wave}

For the vielbein \eqref{d10hstrviel1}, the spin connections are
\bea
\hat{\omega}_{01}&=&-\frac{ e^{-mu}}{2 H_KH_Q^{1/8}}\big(H_K^{(1,0)}\hat{e}^2+ H_K^{(0,1)}\hat{e}^3\big),\quad \hat{\omega}_{0 2}=-\frac{ e^{-mu}H_K^{(1,0)}}{2 H_KH_Q^{1/8}}\Big[\hat{e}^1-\hat{e}^0\big(1+\frac{3 H_K H_Q^{(1,0)}}{4 H_K^{(1,0)} H_Q}\big)\Big],\cr
\hat{\omega}_{0 3}&=&-\frac{ e^{-mu}H_K^{(0,1)}}{2 H_KH_Q^{1/8}}\Big[\hat{e}^1-\hat{e}^0\big(1+\frac{3 H_K H_Q^{(0,1)}}{4 H_K^{(0,1)} H_Q}\big)\Big],\qquad
\hat{\omega}_{0 9}=-\frac{m e^{-m u}}{H_Q^{1/8}}\hat{e}^0,\cr
\hat{\omega}_{1 2}&=&-\frac{ e^{-mu}H_K^{(1,0)}}{2 H_KH_Q^{1/8}}\Big[\hat{e}^0-\hat{e}^1\big(1-\frac{3 H_K H_Q^{(1,0)}}{4 H_K^{(1,0)} H_Q}\big)\Big],\qquad \hat{\omega}_{1 9}=\frac{m e^{-m u}}{H_Q^{1/8}}\hat{e}^1,\cr
\hat{\omega}_{1 3} &=& -\frac{ e^{-mu}H_K^{(0,1)}}{2 H_KH_Q^{1/8}}\Big[\hat{e}^0-\hat{e}^1\big(1-\frac{3 H_K H_Q^{(0,1)}}{4 H_K^{(0,1)} H_Q}\big)\Big],\qquad \hat{\omega}_{2 3}=\frac{ e^{-mu}}{8 H_Q^{9/8}}\big(H_Q^{(0,1)}\hat{e}^2- H_Q^{(1,0)}\hat{e}^3\big),\cr
\hat{\omega}_{2 4} &=&-\frac{e^{-m u} H_Q^{(1,0)}}{8 H_Q^{9/8}}\hat{e}^4,\qquad \hat{\omega}_{4 5}=\frac{2me^{-m u} }{ H_Q^{1/8}}\big(\hat{e}^8-2\sqrt{2}\cot\theta_1\hat{e}^5\big),\cr
\hat{\omega}_{2 5}&=&-\frac{e^{-m u} H_Q^{(1,0)}}{8 H_Q^{9/8}}\hat{e}^5,\qquad \hat{\omega}_{2 6}=-\frac{e^{-m u} H_Q^{(1,0)}}{8 H_Q^{9/8}}\hat{e}^6,\qquad
\hat{\omega}_{2 7}=-\frac{e^{-m u} H_Q^{(1,0)}}{8 H_Q^{9/8}}\hat{e}^7,\cr
\hat{\omega}_{2 8}&=&-\frac{e^{-m u} H_Q^{(1,0)}}{8 H_Q^{9/8}}\hat{e}^8,\qquad
\hat{\omega}_{2 9}=\frac{e^{-m u} }{8 H_Q^{9/8}}\big(8mH_Q\hat{e}^2-H_Q^{(1,0)}\hat{e}^9\big),\qquad \hat{\omega}_{3 4}=-\frac{e^{-m u} H_Q^{(0,1)}}{8 H_Q^{9/8}}\hat{e}^4,\cr
\hat{\omega}_{3 5}&=&-\frac{e^{-m u} H_Q^{(0,1)}}{8 H_Q^{9/8}}\hat{e}^5,\qquad
\hat{\omega}_{3 6}=-\frac{e^{-m u} H_Q^{(0,1)}}{8 H_Q^{9/8}}\hat{e}^6,\qquad \hat{\omega}_{3 7}=-\frac{e^{-m u} H_Q^{(0,1)}}{8 H_Q^{9/8}}\hat{e}^7,\cr
 \hat{\omega}_{3 8}&=&-\frac{e^{-m u} H_Q^{(0,1)}}{8 H_Q^{9/8}}\hat{e}^8,\qquad
\hat{\omega}_{3 9}=\frac{e^{-m u} }{8 H_Q^{9/8}}\big(8mH_Q\hat{e}^3-H_Q^{(0,1)}\hat{e}^9\big),\cr
\hat{\omega}_{4 8}&=&\frac{2me^{-m u} }{ H_Q^{1/8}}\hat{e}^5,\qquad
\hat{\omega}_{4 9}=\frac{me^{-m u} }{ H_Q^{1/8}}\hat{e}^4,\qquad \hat{\omega}_{5 8}=-\frac{2me^{-m u} }{ H_Q^{1/8}}\hat{e}^4,\qquad \hat{\omega}_{5 9}=\frac{me^{-m u} }{ H_Q^{1/8}}\hat{e}^5,\cr
\hat{\omega}_{6 7}&=&\frac{2me^{-m u} }{ H_Q^{1/8}}\big(\hat{e}^8-2\sqrt{2}\cot\theta_2\hat{e}^7\big),\qquad
\hat{\omega}_{6 8}=\frac{2me^{-m u} }{ H_Q^{1/8}}\hat{e}^7,\qquad \hat{\omega}_{6 9}=\frac{me^{-m u} }{ H_Q^{1/8}}\hat{e}^6,\cr
\hat{\omega}_{7 8}&=&-\frac{2me^{-m u} }{ H_Q^{1/8}}\hat{e}^6,\qquad
\hat{\omega}_{7 9}=\frac{me^{-m u} }{ H_Q^{1/8}}\hat{e}^7,\qquad
\hat{\omega}_{8 9}=\frac{me^{-m u} }{ H_Q^{1/8}}\hat{e}^8.
\eea
The form fields are
\bea
\hat{F}_{(2)}&=&\frac{4\sqrt{2}me^{-2mu}}{H_Q^{1/4}}\big(\hat{e}^4\wedge \hat{e}^5-\hat{e}^6\wedge \hat{e}^7\big),\cr
\hat{H}_{(3)}&=&-\frac{e^{-3mu}}{H_Q^{11/8}} \big(H_Q^{(1,0)}\hat{e}^2+H_Q^{(0,1)} \hat{e}^3\big)\wedge\hat{e}^0\wedge \hat{e}^1+\frac{4me^{-3mu}}{H_Q^{3/8}}\big(\hat{e}^4\wedge \hat{e}^5+\hat{e}^6\wedge \hat{e}^7\big)\wedge \hat{e}^8,\cr
d\hat{\phi}&=&-\frac{e^{-mu}}{2H_Q^{9/8}}\big(H_Q^{(1,0)}\hat{e}^2+H_Q^{(0,1)}\hat{e}^3+8mH_Q\hat{e}^9\big)
\eea

\subsection{Heterotic Euclidean instanton}

For the vielbein \eqref{d10hstrviel2}, the spin connections are
\bea
\hat{\omega}_{0 1}&=&-\frac{e^{-m u} (8 H_a+3 r H_a')}{8 r H_a^{11/8}}\hat{e}^1,\qquad \hat{\omega}_{0 2}=-\frac{e^{-m u} (8 H_a+3 r H_a')}{8 r H_a^{11/8}}\hat{e}^2,\cr
\hat{\omega}_{0 3}&=&-\frac{e^{-m u} (8 H_a+3 r H_a')}{8 r H_a^{11/8}}\hat{e}^3,\qquad
\hat{\omega}_{0 4}=\frac{e^{-m u} H_a'}{8 H_a^{11/8}}\hat{e}^4,\qquad
\hat{\omega}_{0 5}=\frac{e^{-m u} H_a'}{8 H_a^{11/8}}\hat{e}^5,\cr
\hat{\omega}_{0 6} &=& \frac{e^{-m u} H_a'}{8 H_a^{11/8}}\hat{e}^6,\qquad \hat{\omega}_{1 2}=-\frac{e^{-m u}}{r H_a^{3/8}}\hat{e}^3,\cr
\hat{\omega}_{0 7}&=&\frac{e^{-m u} H_a'}{8 H_a^{11/8}}\hat{e}^7,\qquad
\hat{\omega}_{0 8}=\frac{e^{-m u} H_a'}{8 H_a^{11/8}}\hat{e}^8\,,\qquad
\hat{\omega}_{0 9}=\frac{e^{-m u}}{8 H_a^{11/8}}\big(8mH_a^{3/2}\hat{e}^0+H_a'\hat{e}^9\big),\cr
\hat{\omega}_{1 3}&=&\frac{e^{-m u}}{r H_a^{3/8}}\hat{e}^2,\quad \hat{\omega}_{1 9}=me^{-m u}H_a^{1/8}\hat{e}^1,\quad
\hat{\omega}_{2 3}=-\frac{e^{-m u}}{r H_a^{3/8}}\hat{e}^1,\quad \hat{\omega}_{2 9}=me^{-m u}H_a^{1/8}\hat{e}^2,\cr
\hat{\omega}_{3 9}&=&e^{-m u}mH_a^{1/8}\hat{e}^3,\quad
\hat{\omega}_{4 5}=2 m e^{-m u}H_a^{1/8}\big(\hat{e}^8-2\sqrt{2}\cot\theta_1\hat{e}^5\big),\quad \hat{\omega}_{4 8}=2 m e^{-m u}H_a^{1/8}\hat{e}^5,\cr
\hat{\omega}_{4 9}&=&m e^{-m u}H_a^{1/8}\hat{e}^4\,,\quad
\hat{\omega}_{5 8}=-2m e^{-m u}H_a^{1/8}\hat{e}^4,\quad \hat{\omega}_{5 9}=m e^{-m u}H_a^{1/8}\hat{e}^5, \cr
\hat{\omega}_{6 7}&=&2 m e^{-m u}H_a^{1/8}\big(\hat{e}^8-2\sqrt{2}\cot\theta_2\hat{e}^7\big),\quad
\hat{\omega}_{6 8}=2m e^{-m u}H_a^{1/8}\hat{e}^7,\quad \hat{\omega}_{6 9}=m e^{-m u}H_a^{1/8}\hat{e}^6,\cr
\hat{\omega}_{7 8}&=&-2m e^{-m u}H_a^{1/8}\hat{e}^6,\quad \hat{\omega}_{7 9}=m e^{-m u}H_a^{1/8}\hat{e}^7,\quad \hat{\omega}_{8 9}=m e^{-m u}H_a^{1/8}\hat{e}^8\,.
\eea
The form fields are
\bea
\hat{F}_{(2)}&=&4\sqrt{2}me^{-2mu}H_a^{1/4}\big(\hat{e}^4\wedge \hat{e}^5-\hat{e}^6\wedge \hat{e}^7\big),\cr
\hat{H}_{(3)}&=&-\frac{e^{-3mu}H_a'}{H_a^{9/8}}\hat{e}^1\wedge \hat{e}^2\wedge \hat{e}^3+4me^{-3mu}H_a^{3/8}\big(\hat{e}^4\wedge \hat{e}^5+\hat{e}^6\wedge \hat{e}^7\big)\wedge \hat{e}^8,\cr
d\hat{\phi}&=&e^{-mu}H_a^{1/8}\Big(\frac{H_a'}{2H_a^{3/2}}\hat{e}^0-4m\hat{e}^9\Big)\,.
\eea

\section{Killing spinors on $S^3$}

The maximally symmetric $S^3$ metric admits maximum number of the Killing spinors. The Killing spinor equations of unit round $S^3$ are given by
\be
\nabla_k \epsilon_\pm  = \pm \ft12 {\rm i} \Gamma_k \epsilon_\pm\,,\qquad k=1,2,3.
\ee
The Killing spinors for spheres written in the nested structures were presented in \cite{Lu:1998nu}. Here, we consider two cases of writing the $S^3$ metric and its vielbein and obtain corresponding Killing spinors. Although the total number of the Killing spinors are the same, they have very different coordinate dependences.

Case one is to write the metric using $SU(2)$-left invariant 1-forms $\sigma_i$, given by \eqref{su21form}. The metric and the natural choice of the vielbein are
\be
ds^2=\ft14 (\sigma_1^2 + \sigma_2^2 + \sigma_3^2)\,,\qquad e^i = \ft12 \sigma_i\,.
\ee
Choosing the $\Gamma$ matrices as the standard Pauli matrices, we find that the two Killing spinors $\epsilon_-$ are simply constant spinors.  The two Killing spinors $\epsilon_+$ are more complicated, depend on all Euler angles:
\be
\epsilon_{+,1} =\left( \begin{array}{c}
\cos \big(\frac{\theta }{2}\big) e^{\frac{1}{2} \mathrm{i} (\varphi +\psi )} \\
\mathrm{i}  \sin \big(\frac{\theta }{2}\big) e^{\frac{1}{2} \mathrm{i} (\varphi -\psi )}
\end{array}\right),\qquad
\epsilon_{+,2} = \left(\begin{array}{c}
\mathrm{i} \sin \big(\frac{\theta }{2}\big) e^{-\frac{1}{2} \mathrm{i} (\varphi -\psi )}\\
 \cos \big(\frac{\theta }{2}\big) e^{-\frac{1}{2} \mathrm{i} (\varphi +\psi)}
\end{array}\right).
\ee
The second case is to write the metric as a $U(1)$ bundle over $S^2$. The metric and the natural choice of vielbein are
\bea
ds^2 &=& \ft14 (d\psi + \cos\theta d\varphi)^2 +\ft14( d\theta^2 + \sin^2\theta\, d\varphi^2)\,,\cr
e^1 &=& \ft12 d\theta\,,\qquad  e^2=\ft12 \sin\theta\,d\varphi\,,\qquad e^3= \ft12 (d\psi + \cos\theta d\varphi)\,.
\eea
The Killing spinor solutions for  $\epsilon_-$ is simpler, depending only on the fibre coordinates
\be
\epsilon_{-,1} = \left( \begin{array}{c}
e^{-\fft12{\rm i} \psi} \\
0\end{array}\right),\qquad
\epsilon_{-,2} = \left(\begin{array}{c}
0\\
 e^{\fft12 {\rm i} \psi}
\end{array}\right).
\ee
The Killing spinors $\epsilon_+$ are more complicated, depending on the base $S^2$ coordinates:
\be
\epsilon_{+,1} =\left(\begin{array}{c}
 {\rm i} e^{-\frac{\mathrm{i} \varphi}{2}} \sin \big(\frac{\theta }{2}\big) \\
 e^{-\frac{\mathrm{i} \varphi}{2}} \cos \big(\frac{\theta }{2}\big)
\end{array}\right),\qquad
\epsilon_{+,2} = \left(\begin{array}{c}
e^{\frac{\mathrm{i} \varphi }{2}} \cos \big(\frac{\theta }{2}\big)\\
\mathrm{i}  e^{\frac{\mathrm{i} \varphi }{2}} \sin \big(\frac{\theta }{2}\big)
\end{array}\right).
\ee

\end{document}